\documentclass[letterpaper]{article}
\usepackage{aaai2026}
\usepackage{times}
\usepackage{helvet}
\usepackage{courier}
\usepackage[hyphens]{url}
\usepackage{graphicx}
\usepackage{natbib}
\usepackage{caption}
\usepackage{booktabs}
\usepackage{tabularx}
\usepackage{array}
\title{Govern, Map, Measure, Absorb: The Legibility Trap\\
in Public Sector Participatory AI Governance}

\author{
    Srijan Pandey\textsuperscript{\rm 1},
    Devansh Saxena\textsuperscript{\rm 1}
}
\affiliations{ \textsuperscript{\rm 1}The Information School, University of Wisconsin--Madison\\ Madison, WI 53706, USA\\ spandey43@wisc.edu, devansh.saxena@wisc.edu}

\begin{document}

\maketitle

\begin{abstract}
Public sector agencies have adopted participatory design as a prominent procedural safeguard proposed against algorithmic harm. Across methods, institutional contexts, and well-intentioned implementations, the participation gap persists. Communities engage, input is documented, and systems are deployed largely unchanged. This paper locates the persistence of that gap in the mechanism institutions use to document participation itself. We develop the concept of the legibility trap, in which institutional mechanisms for documenting participation systematically select for forms of community input that fit institutional categories and filter out the forms that would most constrain state power, including foundational dissent, demands for non-deployment, and what Kelly Oliver calls witnessing to structural conditions of harm. Drawing on James Scott's theory of state simplification, we demonstrate the legibility trap through critical discourse analysis of twenty-two official, community-produced, and public-record documents across three public sector AI deployments: the Allegheny Family Screening Tool, Detroit Project Green Light, and the Community Control Over Police Surveillance (CCOPS) ordinances in Oakland. We identify the structural conditions under which participation partially escapes the trap, and we theorize that repeated non-response constitutes a harm to community subjectivity that AI ethics frameworks have not yet named.
\end{abstract}

\section{Introduction}

Public sector agencies increasingly face organized public resistance to AI infrastructure, from data center siting fights \citep{depillis2026local,Kollar03042026} to algorithmic policing tools, and have responded in divergent ways. Some jurisdictions have tightened procedural requirements for public input. Others have moved to preempt local review altogether. This paper examines a narrower and more persistent problem inside the mechanism most often proposed as the remedy for algorithmic harm, participatory design itself.

Participatory AI governance has become highly effective at documenting community input without allowing it to constrain institutional power. While broader Participatory Design (PD) scholarship has long acknowledged the limits, constraints, and critiques of participatory practice \citep{maas2024beyond,10.1145/3772363.3790077,corbett2023power,birhane2022power}, this paper intervenes in efforts to resist its reduction to documentable metrics, procedural checklists, or a depoliticized design technique, and instead re-centers its origins as a political project grounded in Marxist organizing ideals. Public sector agencies across the United States deploy AI systems that determine whether families receive food assistance, whether children are removed from their homes, and whether individuals are surveilled by police \citep{eubanks2018automating,benjamin2019race}. These systems act on the bodies and lives of people who have no meaningful role in shaping them, particularly those Sara Ahmed describes as ``(dis)oriented'' \citep{ahmed2006orientations}. Community workshops, public hearings, and participation clauses in vendor contracts have become a prominent procedural safeguard proposed against these harms \citep{saxena2025emerging,delgado2023participatory}.

Implementing ``PD as remedy'' has proven ineffective. A recent survey of more than 80 papers indicates that the majority of projects remain consultative, using communities as informants or proxies without restoring agency to at-risk populations \citep{delgado2023participatory}. This recurring pattern suggests that participation is operationalized as a means of improving input quality instead of redistributing decision-making authority, and it stabilizes existing governance arrangements more often than it unsettles them. Prior work documents how participatory processes can legitimate decisions communities were never empowered to contest and fail to address structural problems in machine learning governance. Agency decision-makers may recognize these inadequacies while remaining constrained by institutional power relations, with one public-sector director describing power-sharing as something that ``feels a little foreign'' \citep{birhane2022power,sloane2022participation,kawakami2024studying}. Each diagnosis succeeds analytically while remaining largely descriptive. What is missing is an account of persistence, an explanation of why the gap is continually reproduced despite sustained critique, methodological innovation, and institutional uptake of participatory frameworks. Recent work sharpens this critique by arguing that participation lacks enforceable authority, or ``teeth'' \citep{maas2024beyond}. As long as decisions about stakeholder inclusion remain in developers' hands, participation is arbitrary power dressed as democratic practice. They call for legal-institutional arrangements and alternative ownership models. It does not yet explain why legal-institutional arrangements for participation also fail to produce the accountability they promise. The Community Control Over Police Surveillance (CCOPS) ordinance in Oakland, one of the strongest legal mechanisms for community participation in surveillance technology procurement in the United States, was passed with an ``exigent circumstances'' clause that its own organizers immediately identified as a mechanism for institutional evasion \citep{irani2021oversight}. The problem runs deeper than the absence of legal teeth. It runs through the logic by which institutions make participation legible, thereby reducing PD to a method. Recent research documents how the burden of genuine participatory practice falls on individuals who invest personal time, financial resources, and emotional labor \citep{10.1145/3772363.3790077}. This body of work has documented the participation gap with real precision across scale mismatches \citep{suresh2024participation}, generative AI contexts \citep{10.1145/3772363.3778759}, and procurement constraints \citep{10.1145/3772363.3778739}, but has less often specified the documentation mechanism through which these problems persist together across contexts, methods, and well-intentioned implementations. The mechanism this paper identifies also extends beyond participatory AI governance in the strict sense. A structurally similar pattern appears in workplace AI for feminized labor, where the systemic undervaluing of frontline workers' expertise forms a self-reinforcing loop across design, evaluation, and deployment \citep{doi:10.1177/20539517261424164, saxena2024algorithmic, spektor2025working}. The convergence across two domains suggests that legibility-driven filtering of stakeholder knowledge is a general property of institutional AI governance wherever documentation practices sit between affected people and decisions made about them.

We argue that the persistence of the participatory gap is a structural condition, which we term the \textbf{legibility trap}. Drawing on \citet{scott1998seeing}, we conceptualize legibility as the process through which institutions render social input administrable by simplifying it into standardized and auditable forms. In AI governance settings, participation gains institutional value when it can be documented through categories such as attendance counts, coded feedback, traceable logs, and compliance records \citep{saxena2025emerging}. This proceduralization does not simply record participation but defines its boundaries in advance. Forms of engagement that exceed these categories are systematically excluded from recognition. These exclusions include foundational dissent, refusal, and calls for non-deployment as well as what \citet{oliver2001witnessing} describes as witnessing structural harm, forms of articulation that resist translation into reporting formats. Participation becomes visible to institutions only when it is already shaped to be governable, which excludes expressions that directly contest institutional power. The \textit{legibility trap} is produced by the structural logic of institutional accountability, independent of the good or bad faith of any individual actor. This is why the call for participation ``teeth,'' though necessary, is insufficient \citep{maas2024beyond}. Mandatory legal requirements for community engagement still require that engagement to be documented and verified, which means they still select for forms of participation that can be documented and verified. The same filtering logic operates at the regulatory level and at the workshop level.

Beneath the legibility trap lies a deeper problem the participatory AI literature has not named. Much of the participatory AI literature inherits a liberal-democratic assumption \citep{arnstein1969ladder,schuler1993participatory,delgado2023participatory}. The state was treated as a liberal democratic actor formally accountable to citizens, and communities were treated as holding symmetric civic standing to make governance claims against it. This assumption does not describe the experience of Black communities under facial recognition surveillance in majority-Black cities, families whose poverty is scored as child maltreatment risk, or Dalit communities navigating Aadhaar-linked welfare infrastructure \citep{couldry2019costs,blili2024unsocial,ajunwa2025ai, eubanks2018automating,benjamin2019race}. For these communities, the state's relationship is surveillance, conditionality, and discipline. In that relationship, participation fails to produce change. It actively legitimates harm as consent. The institution can truthfully say it consulted the community, but that consultation becomes a disguise.

There is a third dimension of this failure that the literature has not theorized. Drawing on \citet{oliver2004witnessing}'s distinction between testimony and witnessing, we argue that participatory AI governance processes are structured to receive community testimony in the form of documented preferences and thematic concerns. These processes do not have the capacity to receive community witnessing, which involves bearing witness to conditions of structural harm that exceed design parameters and institutional categories of recognition. Participation processes that collect testimony without producing substantive response generate more than inefficiency or wasted labor. They produce a patterned absence of address. In \citet{oliver2004witnessing}, the address response relation is central to the formation of subjectivity under conditions of oppression. When this relation is repeatedly broken, participation becomes a site where agency is eroded through sustained non-response. This produces a form of harm that is distinct from policy failure in which systems are deployed despite opposition. It is a harm that operates at the level of subject formation itself. The experience described by Tawana Petty after years of organizing against Detroit's facial recognition system, where she states she is ``beyond exhausted,'' illustrates how prolonged participation without institutional uptake exhausts the conditions of political address and undermines subjectivity \citep{wdet2020detroit, bridgedetroit2021ciogs}. This paper is organized around three questions: 1) how do institutional documentation mechanisms transform community knowledge in ways that undermine its capacity to constrain state power, 2) what structural conditions differentiate cases where participation produced binding accountability from cases where it produced documented but inconsequential engagement, and 3) what does it mean to treat participation as a political project oriented toward power redistribution instead of a design method oriented toward input collection.

We address these questions through critical discourse analysis (CDA) of official participatory documents across three public sector AI deployments: the Allegheny Family Screening Tool, Detroit Project Green Light, and CCOPS Oakland. Each case surfaces a distinct articulation of the \textit{legibility trap}, including foreclosure at the level of procurement, absorption of participation without substantive consequence, and the limited conditions under which participation partially escapes institutional capture. Across cases, we show that the legibility trap is not an unintended effect of implementation. It is embedded in the institutional language through which participation is recorded and evaluated. That language presumes in advance what counts as participation and what can be recognized as its outcome. Oakland's partial escape from the trap does not falsify the structural argument. Instead, it clarifies the conditions under which the trap can be displaced onto a different institutional location, which is itself a structural finding. This paper is also an argument about data. Public agencies do not simply consult communities and record what happened. They convert participation into administrative data, meeting counts, coded feedback, procurement artifacts, compliance reports, and audit logs, through which community input becomes something an institution can process. Those records do not preserve community voice in a neutral form. They define in advance what forms of voice can become institutionally actionable. The legibility trap is therefore a datafication problem inside democratic governance. Community participation becomes governable exactly when it has been rendered into categories that no longer threaten institutional authority.

\section{Related Work}

\subsection{The Political Origins of Participatory Design}

Participatory design originated as a political project for redistributing power over the technologies governing workers' lives \citep{ehn1988work,bjerknes1987computers,10.1108/eb055278}. The UTOPIA project produced a legal agreement backed by trade union authority; the accountability mechanism that made participation meaningful was the union, countervailing power structurally independent of the institution being governed \citep{bratteteig2018pd}. Arnstein's ladder, developed during urban renewal and the civil rights movement, was equally a political document \citep{arnstein1969ladder}. The argument that community control requires authority communities hold independently of the institution was largely lost when the ladder became a design rubric in HCI \citep{delgado2023participatory,corbett2023power}.

A parallel tradition developed in the Global South: participatory action research (PAR) in Colombia was developed explicitly against extractive research and for liberation \citep{falsborda1987}, and conscientization was theorized as structural transformation of understanding, not preference collection \citep{freire1970}. Both traditions treated participation as a political practice \citep{mohamedecolonial2020,suresh2022feminist}. The HCI literature's participatory turn encodes a Eurocentric genealogy by citing Scandinavia while omitting this broader PAR tradition. When participatory design migrated into North American HCI, a political project was translated into a design methodology \citep{schuler1993participatory}. Costanza-Chock names this as the loss of design justice \citep{costanza-chock2020design}, and community-based collaborative design frequently extracts knowledge without returning value \citep{harrington2019deconstructing}. The depoliticization was the condition under which participatory design became institutionally legible as a method.

\subsection{The Emerging Critique: Seeing the Problem, Staying Inside It}

The participatory AI literature produces consistent diagnoses of the participation gap \citep{delgado2023participatory,corbett2023power,birhane2022power, sloane2022participation}. What remains underdeveloped is a structural explanation for its persistence. The most comprehensive survey identifies gaps between consultative and transformative participation, proxy engagement, and uneven power distribution \citep{delgado2023participatory}. Earlier work by the same authors shows that adding stakeholders does not alter structural conditions of decision-making \citep{delgado2021stakeholder}. Co-optation risks and the structural limits of participation in ML governance are documented across multiple lines of critique \citep{birhane2022power,sloane2022participation,palacin2020design}. Across clusters of scholarship, the gap is named and then addressed through proposals for improved institutional mechanisms, pseudo-participation is distinguished from genuine engagement, and remedies are proposed for improving participatory processes \citep{feffer2023preference,robertson2020limits}. Empirical work shows institutional power structures shaping AI systems before engagement begins \citep{kawakami2024studying,johnson2025legacy}, community interventions challenging system premises that institutional processes lack mechanisms to incorporate \citep{haque2024asking,kuo2023understanding,martin2020participatory}, and communities articulating alternative futures that exceed institutional categories of response \citep{10.1145/3686905, stapleton2022imagining}. Even enforceable participation requirements can remain inside institutional categories of auditable engagement, preserving the structural filtering of what counts as governance input \citep{maas2024beyond}. This body of work has documented these failures with real precision, but has less often specified the documentation mechanism through which they persist together across contexts, methods, and well-intentioned implementations.

\subsection{Practices, Frameworks, and the Recognition Structure}

Current responses to participation washing formalize participation as an accountable object through audit log systems, representation metrics, and lifecycle governance models \citep{tabassi2023ai,gourlet2024reclaiming}. Scott's analysis of state simplification clarifies the consequence \citep{scott1998seeing}: more refined measurement systems more efficiently filter forms of participation that cannot be translated into institutional categories \citep{brilmyer2023terms}. Oliver's account of recognition clarifies the underlying epistemic structure \citep{oliver2001witnessing}. The institution retains the authority to grant recognition while communities remain dependent on that grant. Across the literature, participation is framed as a problem of improving recognition instead of transforming the conditions that define recognition \citep{corbett2023power}.

Political economic accounts locate asymmetry in ownership of infrastructure, data, and value capture \citep{maas2024beyond,ajunwa2023artificial}. In child welfare, homelessness services, and public education, labor is absorbed into data infrastructures and disappears from institutional representation, with care work and interpretive labor converted into categories compatible with reporting systems \citep{saxena2022unpacking,moon2025datafication,puig2015matters, 10.1145/3805689.3812281}. The distinction between testimony and witnessing is central to this transformation \citep{oliver2004witnessing}. Testimony aligns with institutional legibility; witnessing exceeds it by articulating structural harm that cannot be captured within administrative categories. Participatory processes retain testimony while filtering witnessing, which eventually leads to subjectivity harm, manufactured through prolonged institutional non-response. The affective and labor burdens of that non-response produce exhaustion documented across contexts \citep{10.1145/3772363.3790077,fox2017intersectional, doi:10.1177/20539517261424164}. The legibility trap provides a structural account of this persistence through Scott's theory of administrative simplification and Oliver's theory of recognition and witnessing. The next section develops this account empirically.

\section{Method}

\paragraph{Approach.}

This paper uses critical discourse analysis (CDA) of publicly available governance documents, triangulated with cross-case analysis of three public sector AI deployments. CDA follows directly from the theoretical argument. Methodologically, the paper combines CDA with critical case comparison. CDA identifies the textual mechanisms through which participation is made institutionally legible. Cross-case comparison identifies the institutional conditions under which those mechanisms are intensified, displaced, or partially interrupted. The claim is fundamentally discursive. Institutional documentation mechanisms transform community knowledge in specific ways visible in how institutions write participation into existence in official language \citep{fairclough1995critical,bowker1999sorting,dIgnazio2020data}. Interview-based methods capture what practitioners say about participation. Document analysis shows what institutions construct participation as having produced. The gap between those two accounts is where the legibility trap operates \citep{cooper2024fitting,gourlet2024reclaiming}. This choice follows Fairclough's critical discourse analysis tradition, which reads textual features as traces of the social relations and power structures that produced them \citep{fairclough1995critical}. It also follows a broader humanistic HCI tradition of combining discourse analysis with interpretive close reading of institutional and administrative texts \citep{bardzell2015humanistic}. Applied here, that means treating the grammatical subject of a sentence, the placement of a quotation, and the presence or absence of a category as data in their own right. A survey of practitioner interviews could establish that officials believe they engaged communities meaningfully. It could not establish what the official record does with that engagement once it is written down, and that is the paper's actual claim. Analyzing the conditions under which community knowledge is or is not received as governance input requires attending to the official record itself, not only to practitioner accounts of it \citep{dIgnazio2020data}. The case study component follows Flyvbjerg's critical case logic, in which each case was selected for analytic variation across different registers of the legibility trap \citep{flyvbjerg2006five}. Allegheny shows the trap operating at procurement and design, before any community engagement process begins. Detroit shows the trap in a fully documented formal participation process that produced no change. Oakland shows the structural conditions under which participation partially escaped the trap.

\paragraph{Corpus Construction.}

The corpus was constructed in three stages. First, all publicly available official documents produced by each institution in connection with participatory processes surrounding each AI deployment were identified. Document types include procurement documents, development reports, ethical reviews, process and impact evaluations, policy directives, hearing records, ordinance texts, annual accountability reports, and official public statements. Second, inclusion criteria required documents to be produced or commissioned by the public institution and to bear directly on the participatory processes surrounding the deployment. Third, one community-produced document per case that entered into an official public record was included. These documents provide the analytical baseline against which institutional language can be compared, making visible what community organizations themselves said versus what the official record constructs them as having said. This produced twenty-two official, community-produced, and public-record documents across three cases. Table~\ref{tab:corpus} presents the full corpus. All documents are publicly available and were confirmed through the authors' independent search.A supplementary corpus registry listing all 22 documents with document 
type, source institution, primary and secondary URLs, and access dates 
is available at \url{https://doi.org/10.5281/zenodo.21833385} 
\citep{pandey2026appendix}.

\begin{table}[!t]
\centering
\setlength{\tabcolsep}{3.5pt}

\footnotesize

\begin{tabularx}{\linewidth}{
>{\centering\arraybackslash}p{0.48cm}
>{\raggedright\arraybackslash}p{1.20cm}
>{\raggedright\arraybackslash}X
>{\raggedright\arraybackslash}p{1.05cm}
>{\centering\arraybackslash}p{0.62cm}}

\toprule

\textbf{ID} & \textbf{Case} & \textbf{Document} & \textbf{Type} & \textbf{Yr} \\

\midrule

A1 & Allegheny & RFP: Decision Support Tools & Proc. & 2014 \\

A2 & Allegheny & Developing Predictive Models (Vaithianathan et al.) & Dev. Rpt. & 2017 \\

A3 & Allegheny & Ethical Analysis and DHS Response & Ethics & 2017 \\

A4 & Allegheny & Process Evaluation (Hornby Zeller) & Proc. Eval. & 2018 \\

A5 & Allegheny & AFST FAQ & FAQ & 2017 \\

A6 & Allegheny & AFST FAQ Updated & FAQ & 2019 \\

A7 & Allegheny & Impact Evaluation (Goldhaber-Fiebert et al.) & Impact Eval & 2019 \\

A8 & Allegheny & Evaluation Findings on Predictive Risk Models & Policy & 2024 \\

\midrule

D1 & Detroit & Project Green Light Agreements Page & Web & 2016 \\

D2 & Detroit & DataWorks Plus Contract (No.\ 6000801) & Proc. & 2017 \\

D3 & Detroit & DPD Facial Recognition Policy, Directive 307.5 & Policy & 2019 \\

D4 & Detroit & DPD Draft Facial Recognition Policy (Directive 307.5, July 2019 draft) & Policy Rev & 2019 \\

D5 & Detroit & WDET: City Council Vote on Facial Recognition Contract & Pub. Rec. & 2020 \\

D6 & Detroit & Detroit Community Technology Project Summary & Comm. & 2019 \\

D7 & Detroit & CIOGS Ordinance & Ord. & 2021 \\

\midrule

O1 & Oakland & PAC Establishing Ordinance 13349 C.M.S. & Ord. & 2016 \\

O2 & Oakland & Surveillance Equipment Transparency Ordinance (O.M.C.\ 9.64) & Ord. & 2021 \\

O3 & Oakland & Implementation Guidance: Oakland Privacy Principles & Comm. & 2019 \\

O4 & Oakland & OPD ALPR Annual Report & Ann. Rpt. & 2020 \\

O5 & Oakland & PAC Meeting Minutes and Agenda Packets & Mtg. Rec. & 2020--21 \\

O6 & Oakland & Secure Justice Lawsuit Petition \& Public Reporting on OPD Non-Compliance & Legal & 2021 \\

O7 & Oakland & Annual Surveillance Report to City Council & Ann. Rpt. & 2024 \\

\bottomrule

\end{tabularx}
\caption{Document Corpus for Critical Discourse Analysis}
\label{tab:corpus}
\end{table}

\paragraph{CDA Coding Scheme.}

Five coding categories, derived from the theoretical framework, were applied consistently across the corpus. \textit{Voice representation} captures whether community voice appears as direct quotation, paraphrase, aggregate statistic, or is absent from the document entirely, tracking who the official record constructs as a speaking subject and who it constructs as a data subject or beneficiary. \textit{Input language} captures the specific words used to construct community contribution, from feedback and concerns to opposition, refusal, and absence, tracking the transformation of community input as it moves through institutional processing. \textit{Decision language} captures the words used to construct institutional authority and decision rationale, including evidence-based, data-driven, and operational necessity, tracking whether institutional decisions are constructed as accountable to community input or as technically autonomous. \textit{Structural position} captures where community voice appears in the document relative to institutional analysis. Community voice in appendices, footnotes, or acknowledgment sections constructs it as peripheral. Community voice as a precondition for institutional action constructs it as constitutive. \textit{Absence markers} capture what communities said in contemporaneous reporting or advocacy documents that does not appear in the official document, since absence is itself a discursive act. What the official record excludes is as analytically significant as what it includes \citep{bowker1999sorting,dIgnazio2020data,brilmyer2023terms}. Absence markers were established only where community-produced documents, public testimony, or contemporaneous advocacy records in the corpus provided explicit evidence of claims that did not appear in the corresponding official institutional document, a bounded cross-reference operation rather than open-ended inference.

\paragraph{A~Worked Example.} An example from the corpus shows the five categories applied together, using a sentence already discussed in the Allegheny findings below. Document A6 states that community engagement has been a priority for the county throughout the project, and that the county sought input through six project-specific meetings. Under voice representation, the sentence constructs the county as the speaking subject; no individual or affected family appears as a named speaker. Under input language, community contribution is framed through the word priority and a count of meetings held. Both terms describe institutional process. Neither describes what was said inside those meetings. Under decision language, this particular sentence carries no decision rationale. Elsewhere in the document, decision language attaches to accuracy and predictive validity. It does not attach to the content of community concerns. Under structural position, the sentence sits prominently in the FAQ document, which gives community engagement the appearance of institutional centrality even as its content remains procedural. Under absence markers, the sentence records that meetings occurred without recording what was said in them, a gap made visible only by reading document A4's process evaluation alongside it. The five categories, applied to a single sentence, show how a document can register participation as present while withholding any trace of its content.

\paragraph{Analytic Process and Reflexivity.}

The first author conducted all primary coding chronologically within each case, tracing how institutional language evolved from procurement through post-deployment evaluation. An initial codebook was developed inductively from the first ten documents and applied deductively to the remaining twelve. The second author independently coded all twenty-two documents across the three cases. Following the humanistic HCI tradition, which holds that rigor in interpretive work is demonstrated through theoretical coherence, thick description, and deliberate critical scrutiny rather than through inter-rater reliability coefficients \citep{bardzell2015humanistic, mcdonald2019reliability}, codes were compared through deliberation. Disagreements clustered primarily around the boundary between input language and absence markers and were resolved by returning to the relevant community-produced source document alongside the institutional record. The corpus was also coded for evidence complicating the legibility trap argument, including instances where community voice appears constitutively in institutional language. This was the case in Oakland's O3, and the finding is discussed accordingly. Situated knowledge frames this analytic stance as partial and located \citep{haraway1988situated}. Researcher positionality is treated as a condition of knowledge production, not a limitation to be disclosed and set aside \citep{dIgnazio2020data}.

\section*{Positionality} 

Both authors are researchers in human-computer interaction and science and technology studies whose work focuses on algorithmic governance, participatory design, and critical data studies. Neither author was employed by or formally affiliated with any of the three institutions examined in this study at the time of data collection or analysis. The first author's research on civic AI governance and critical feminist approaches informed the analysis of how public institutions represent community knowledge, distribute authority, and render participation administratively legible. The second author's research on public sector AI governance and participatory design informed our attention to how institutional processes structure opportunities for community participation. These research backgrounds informed our attention to what institutional language excludes and to what community testimony cannot express within institutional categories. Discrepancies in coding, particularly around absence markers, were resolved through deliberate reflection on these positioned readings.

\section{Findings and Discussion}

\subsection{Case 1: Allegheny County AFST: Foreclosure at Procurement}

The Allegheny Family Screening Tool (AFST) was deployed in August 2016 to score families for child maltreatment risk when a call alleging maltreatment is received by the county child welfare hotline. The deployment followed a four-year process that included competitive procurement, an external ethical review, a process evaluation, and an impact evaluation. Allegheny County DHS described this process as a model of responsible AI governance. The official documents tell a more precise story, one that extends what participatory governance scholars have documented about proxy substitution and technical neutralization into the language of the official record itself.

\paragraph{CDA Finding 1. The Proxy Substitution.}

Document A6 states: ``Community engagement has been a priority for the County throughout the project. The County sought input from the community through various meetings, including six project-specific meetings. Three were held at early stages of the project to collect feedback from key external stakeholders.'' The sentence performs a substitution the rest of the document confirms. The external stakeholders named in the process evaluation (A4) are DHS staff, caseworkers, supervisors, agency leadership, community service providers, advocacy groups, foundations, and a family court judge. Families whose data enters the AFST, whose poverty is algorithmically processed, do not appear as participants in any document in this twenty-two-document corpus. They appear as data subjects. The word ``community'' constructs service providers as a community. The proxy substitution is not named. It is performed through vocabulary.

Prior work documents agency reliance on professional intermediaries and identifies proxy engagement as a core feature of consultative participation \citep{kawakami2024studying,delgado2023participatory}. CDA here shows how that substitution appears inside the official document as a categorical construction that no single actor produces, and that remains invisible as substitution \citep{agnew2024illusion}.

Read alongside A6, the process evaluation (A4) specifies what community engagement was constructed as doing. Community meetings discussed ``whether or not the tool would collect or share any new data regarding families. Presenters explained that the tool only leverages data that are already collected and owned by the County.'' The content of engagement was a question about data ownership. The poverty-as-pathology challenge that \citet{brown2019toward} document in their participatory research with affected families does not appear in any official document as a design input raised, received, considered, or rejected. The challenge was never legible as a design input. The categories of the engagement apparatus could not receive it \citep{eubanks2018automating,chang2022expansiveparticipatoryaisupporting}. What \citet{haque2024asking} call foundational dissent, the questioning of the system's premise, was filtered before participation began.

\paragraph{CDA Finding 2. Technical Neutralization.}

Document A3, the ethical review by Dare and Gambrill, finds the AFST ethically appropriate, concluding that ``it is hard to conceive of an ethical argument against use of the most accurate predictive instrument'' and that there are ``significant ethical issues in not using the most accurate risk prediction measure'' \citep{a3dare2017}. The ethical analysis operates entirely within a framework of predictive accuracy. Whether poverty indicators should serve as child maltreatment risk proxies is not raised as an ethical question.Document A8, the 2024 DHS evaluation, evaluates the system in terms 
of classification accuracy and racial disparities in screen-in rates. An independent mixed-methods analysis found that the algorithm alone produced a 20\% disparity in screen-in rates between Black and white children, while workers reduced that disparity to 9\% through holistic risk assessments and adjustments for the model's limitations \citep{10.1145/3491102.3501831}. The observed reduction therefore reflects worker discretion and intervention rather than the predictive score operating independently. Whether using poverty as risk is itself a structural harm is not a question the evaluation apparatus can ask, because the apparatus measures accuracy and disparity reduction, not the core premise of the system \citep{brown2019toward,stapleton2022imagining}.

Ethical evaluation must attend to the relational structures of power encoded in algorithmic systems, not to technical performance metrics alone \citep{birhane2021relational}. The AFST ethical review shows how a political decision is converted into a technical question through the language of the official document \citep{martin2020participatory}: whether poverty is risk becomes whether the data has predictive validity. Once this conversion occurs, the community challenge has no category in the official record to inhabit \citep{robertson2020limits}.

\subsection{Case 2: Detroit Project Green Light: Absorption Without Consequence}

Detroit Police Department (DPD) integrated facial recognition technology into Project Green Light beginning in 2017 with a contract with DataWorks Plus (Contract No.\ 6000801) signed without public input or council approval. By 2020, a sustained coalition of civil rights organizations had organized opposition through multiple City Council hearings. The official record of that opposition is the richest document set in the corpus for demonstrating the legibility trap in a fully operational formal participation process.

\paragraph{CDA Finding 1. Witnessing Transformed to Documented Opposition.}

Document D5, WDET's public reporting of the September 29, 2020 Detroit City Council vote and public testimony, contains Tawana Petty's public comment: ``We have leveraged facial recognition surveillance against Black residents despite the fact that the technology is racially biased and being banned in predominately white cities''\citep{wdet2020detroit}. This statement is not a complaint about design parameters. It is a statement about the racial structure of surveillance in a majority-Black city, about the asymmetry between where the technology is banned and where it is deployed, about what that asymmetry reveals about the value of Black lives in Detroit.

What Petty was doing in Document D5 is precisely what participatory governance researchers have been unable to theorize: the act of bearing witness to structural conditions of harm that institutional categories cannot receive as governance input. Across public-sector AI domains, affected communities often redirect participation toward whether a system should exist at all or articulate needs outside the scope of what the system is designed to address. CDA here shows a similar transformation in the official record of a single meeting \citep{haque2024asking,kuo2023understanding}. The council deliberation does not engage Petty's statement.

Councilmember James Tate's vote rationale states: ``They want any and all tools that are possibly available to ensure that justice is brought about'' \citep{wdet2020detroit}. The subject of Tate's statement is victims of crime. The subject of Petty's statement is Black residents under surveillance. The deliberation reclassifies the opposition from a structural critique about racial surveillance to a competing preference about public safety tools. Petty's testimony enters the record as opposition and exits the deliberation as a preference to be outweighed. The transformation is complete in the language of the institutional record \citep{agnew2023resistance}. The record shows community speech entering the process while its structural force is converted into a preference to be outweighed.

At the level of governance language, the council record shows participation occurring while leaving no category for the structural harm that the testimony documented \citep{10.1145/3686905}.

\paragraph{CDA Finding 2. Loophole Inscription in the CIOGS Ordinance.}

Document D7, the Community Input Over Government Surveillance (CIOGS) ordinance passed in 2021, is the product of counter-organizing that went outside the standard participation process. The CIOGS ordinance was introduced by Detroit City Council in direct response to years of coalition organizing against the DataWorks Plus facial recognition contract, and represents one of the first binding legislative mechanisms in Detroit requiring community input over surveillance technology acquisition. Unlike the standard procurement process through which the Detroit Police Department (DPD) had acquired the DataWorks contract without council approval, CIOGS mandates a public hearing and council vote before any new surveillance technology is deployed. Its language shows the legibility trap operating at the legislative level. The CIOGS ordinance includes an ``exigent circumstances'' clause allowing DPD to continue or expand surveillance use without council approval in emergencies, with the definition of exigent circumstances left to DPD's determination and the approval process moved to closed session. Tawana Petty's response was immediate. In public reporting on the ordinance, Petty stated, ``A policy that leaves `exigent circumstances' to be defined by the DPD, approved in closed session by Detroit City Council, offers consistent opportunities to negate the policy under repeat extensions, I see as a predictable slippery slope” \citep{bridgedetroit2021ciogs,irani2021oversight}.

The drafting process made the ordinance legible as governance by making it administratively workable for DPD, which required preserving operational discretion. The loophole follows directly from that requirement. This is the legibility trap at the level of legislation. The community demand for a \textit{binding} no was converted into a \textit{documented-but-bypassable} constraint. Even legal mechanisms intended to give participation enforceable authority can preserve discretion through administrative exceptions \citep{maas2024beyond,southerland2023master}.

Petty's documented statement that she was ``beyond exhausted'' is the empirical indicator of what the theoretical framework predicts: the accumulated effect of repeated address without response on the capacity for sustained agency \citep{10.1145/3613904.3641898}. Years of organizing produced a contract renewal, a policy with a loophole, and a council vote documented as community input received. The labor of contestation falls entirely outside the accountability structure the documents construct \citep{10.1145/3772363.3790077}.

\subsection{Case 3: CCOPS Oakland: Structural Conditions for Partial Escape}

Oakland's Privacy Advisory Commission (PAC) was established by Ordinance 13349 C.M.S. in January 2016 following community organizing against the Port of Oakland Domain Awareness Center. The Surveillance Equipment Transparency Ordinance (O.M.C.\ 9.64), passed subsequently and amended in 2021, requires affirmative City Council approval before any surveillance technology acquisition, with PAC review as a precondition. By 2021, twenty-five jurisdictions had passed some version of CCOPS legislation \citep{irani2021oversight}.

\paragraph{CDA Finding 1. Mandatory Language Versus Advisory Language.}

Placing Document A6 (Allegheny FAQ) and Document O2 (Oakland Surveillance Ordinance) side by side produces the clearest textual demonstration of what the legibility trap looks like from each side. In Allegheny's document A6, the text reads, ``Community engagement has been a priority for the County throughout the project.'' The grammatical subject is the County. Community engagement is the County's production. The sentence positions the institution as the active agent and participation as the institutional output. In Oakland's document O2, the ordinance provides that the City Council shall only approve an action after first considering the recommendation of the Privacy Advisory Commission and determining that ``the benefits to the community of the surveillance technology outweigh the costs'' and that civil liberties and civil rights are safeguarded (O2, \citealt{oaklandmc964}). The grammatical structure constrains the City Council. The Council is subject to a legal obligation whose precondition is community consultation. In Oakland's ordinance, participation functions as the institution's precondition. In Allegheny's FAQ, participation functions as the institution's output. The difference sits in institutional architecture, specifically in the legal position participation occupies relative to institutional action \citep{corbett2023power,delgado2022legal}. Legacy procurement structures and agency power relations often determine what participation can achieve before a workshop begins. Oakland inverted this sequence by making community consultation a structural precondition for institutional action \citep{johnson2025legacy,kawakami2024studying}. Community consultation functions there as the structural precondition for institutional action, a different role than the documented record of institutional process it plays elsewhere. The legal architecture makes the difference visible in a single sentence.

Document O3, the Oakland Privacy Coalition's Implementation Guidance, states that the Accountability Principle ``acknowledges that accountability requires more than serving as a passive receptacle for feedback.'' This document is community-produced and was incorporated into the city's official Privacy Principles. The contrast with A6 is a contrast between community-authored accountability language and institution-authored engagement language within the same governance framework. Where Allegheny's FAQ positions the Department of Human Services (DHS) as the subject of community engagement, Oakland's community document names passivity as the failure mode and positions the institution as structurally obligated to respond.

\paragraph{CDA Finding 2. The Residual Trap.}

Document O6, the Secure Justice lawsuit petition and contemporaneous public reporting on OPD non-compliance with Oakland's surveillance ordinance, shows the limits of even the strongest accountability mechanism in the corpus. The lawsuit contended that the City Attorney's Office and OPD violated Oakland's surveillance ordinance by not providing documents the PAC had requested. The PAC oversight model was failing because nine civilian volunteers had no full-time staff to gather information from city departments and hold them accountable. Structural conditions for partial escape from the legibility trap are necessary but not sufficient. The trap reproduces itself at each level of accountability sophistication, as Scott's account of administrative simplification predicts. More sophisticated accountability mechanisms more efficiently filter the forms of engagement they cannot categorize \citep{scott1998seeing,bratteteig2018pd}.

What Oakland shows is that the trap can be moved from whether institutions will acknowledge community opposition to whether they will comply with the law. The community organization that produced the PAC did so through years of political organizing that built legal authority outside the institution being governed \citep{ehn1988work,10.1108/eb055278,falsborda1987}.

\subsection{Cross-Case Synthesis}

Three findings hold across all three cases and constitute the paper's core empirical contribution. 

\textit{Finding 1.} The forms of community input official documents record are precisely the forms that do not constrain institutional action. Allegheny records data security preferences, Detroit records opposition and a vote that overrode it, Oakland records the number of acquisitions reviewed. The poverty-as-pathology challenge, the demand for non-deployment, and the PAC chair lawsuit naming institutional non-compliance, the inputs that would have most constrained institutional power, are systematically absent from every official record.

\textit{Finding 2.} The transformation of community knowledge happens in the official document itself. The proxy substitution in A6, the reclassification of witnessing as opposition in D5, and the loophole inscription in D7 are discursive events in institutional language. Interview-based methods capture what practitioners describe as having happened, document analysis shows what the official record constructs as having happened \citep{dIgnazio2020data,gourlet2024reclaiming,brilmyer2023terms}. The documents function as part of the participation process itself, making it legible for institutional accountability.

\textit{Finding 3.} The cases where community authority produced genuine constraint share structural conditions absent from the others. Oakland partially escaped the legibility trap through four conditions: a binding no inscribed in ordinance language before any specific participation event, an accountability body external to and legally independent of the governed institution, burden of proof sitting with the institution, and ongoing annual accountability. All four conditions were produced by community political organizing that built legal authority outside the institution being governed \citep{ehn1988work,10.1108/eb055278,falsborda1987}. This mirrors the structural condition the Scandinavian PD tradition built its accountability on: worker authority in the UTOPIA project was grounded in trade union power that existed outside and independently of the institution being governed \citep{ehn1988work,bratteteig2018pd}. Oakland's partial escape suggests that what made early PD meaningful was the countervailing power its methods were embedded in. Table~\ref{tab:crosscase} places the three cases in structured comparison. The additional column, what was filtered, makes the absence markers category of the CDA coding scheme visible at the cross-case level. It shows that the forms of community input systematically absent from official records are not random omissions. They are the forms of foundational challenge that each institutional documentation apparatus was architecturally unable to receive. The column pattern is the legibility trap. Table~\ref{tab:crosscase} operationalizes the five coding categories, \textit{participation form} (voice representation), \textit{what was documented} (input and decision language), \textit{what was filtered} (absence markers), and \textit{accountability mechanism} (structural position).

\begin{table}[!t]
\centering
\footnotesize
\setlength{\tabcolsep}{2.5pt}
\begin{tabularx}{\linewidth}{
>{\raggedright\arraybackslash}p{1.35cm}
>{\raggedright\arraybackslash}X
>{\raggedright\arraybackslash}X
>{\raggedright\arraybackslash}X
>{\raggedright\arraybackslash}X} \toprule \textbf{Case} & \textbf{Participation Form} & \textbf{What Was Documented} & \textbf{What Was Filtered} & \textbf{Accountability Mechanism} \\ \midrule Allegheny AFST & Stakeholder meetings with service providers, affected families absent & Data security concerns, technical validation process & Poverty-as-risk challenge, premise of the system & Researchers report to DHS, community has no standing \\ \midrule Detroit Project Green Light & Public hearings, City Council testimony, sustained coalition organizing & Opposition to contract renewal, council vote to extend contract & Non-deployment demand, anti-Black surveillance structure & Advisory board inside governed institution, exigent circumstances loophole \\ \midrule Oakland CCOPS & Council approval required by ordinance, PAC review as precondition & Community authority as constitutive precondition, annual compliance record & PAC document requests refused by OPD & PAC external to OPD, legally binding affirmative approval required \\ \bottomrule \end{tabularx}
\caption{Cross-Case Comparison: Participation Form, Documentation, Outcome, and Accountability Mechanism}
\label{tab:crosscase}
\end{table}

\section{Conclusion}

This paper named the \textit{legibility trap}, demonstrated it empirically across three cases and twenty-two official, community-produced, and public-record documents, and identified the structural conditions under which participation partially escapes it. The argument proceeds in three steps. CDA of these publicly available governance documents reveals that institutional documentation mechanisms systematically select for community input that fits institutional categories and filter out the forms that would most constrain state power, a claim demonstrated in the documents themselves. The cross-case comparison shows that structural conditions differentiating genuine accountability from documented engagement are architectural: where participation produces binding constraint, it does so through a legally inscribed refusal mechanism, an accountability body independent of the governed institution, the burden of proof sitting with the institution, and ongoing compliance obligations. And participation processes that repeatedly collect community testimony without substantive response break the address-response relation through which people sustain political agency, producing a harm at the level of subject formation distinct from adverse policy outcomes. The three cases point toward a more compact statement of the mechanism, one that can travel beyond this particular set of deployments. \textbf{The legibility trap reproduces wherever three conditions hold together:} 1) Institutional accountability depends on documentation, so an agency can only account for what it has recorded \citep{scott1998seeing}, 2) the categories through which participation will be documented are fixed before any specific engagement event occurs, set in advance through procurement language, ethical review templates, or reporting formats, and 3) the forms of address most capable of constraining institutional power, foundational dissent, refusal, and what \citet{oliver2001witnessing,oliver2004witnessing} calls witnessing to structural harm, are precisely the forms that exceed those pre-set categories. When all three conditions hold, more sincere facilitation or more careful methodology cannot open a path out, because the categories that would need to change are set before the workshop begins.

Stated this way, the \textit{legibility trap} gains diagnostic reach past the three cases presented here. Oakland's partial escape becomes legible under the same formulation. Community consultation in Oakland moved from a documented output of institutional process to a legal precondition for institutional action. That is a change specifically to the second condition. An institution that tries to escape the trap by improving consultation records, refining representation metrics, or expanding audit logs is working on the first condition while leaving the second untouched. Scott's account of administrative simplification predicts what follows from that choice. Documentation grows more sophisticated and filters the same way, at a higher resolution.

A predictable counter-argument holds that the three cases are too few and too U.S.-specific to support a structural claim. The response is methodological and theoretical simultaneously. Critical case logic requires that cases be selected for analytic variation across the mechanism being theorized, a different standard than representativeness \citep{flyvbjerg2006five}. Allegheny, Detroit, and Oakland vary across institutional register, method sophistication, and outcome, and the same structural dynamic appears in all three. The theoretical claim rests on Scott's account of state legibility as a general feature of how institutions administer complexity \citep{scott1998seeing}, and the cases provide the empirical demonstration that the mechanism operates in the specific domain of participatory AI governance. A single counter-case where a well-designed co-design method produced genuine community accountability over a deployed public sector AI system, without external legal authority, would falsify the argument. We found no such counter-case in the literature reviewed here \citep{delgado2023participatory,kawakami2024studying,johnson2025legacy}. Three contributions follow from this analysis.

The first is the \textit{legibility trap} as a named structural concept. The field has documented the participation gap with empirical rigor \citep{delgado2023participatory,birhane2022power,sloane2022participation,kawakami2024studying,johnson2025legacy}. This paper adds a structural account of why it persists. Scott's theory of state simplification provides that account, and the CDA methodology operationalizes it at the textual level. The five coding categories---\textit{voice representation}, \textit{input language}, \textit{decision language}, \textit{structural position}, and \textit{absence markers}---constitute a transferable methodological contribution applicable to EU AI Act impact assessments, NIST AI RMF governance reports, and emerging state-level AI accountability documents. 

The second is the theorization of participation failure as subjectivity harm. The AI ethics field has developed frameworks for outcome harm, discriminatory decisions, biased outputs, disparate impacts \citep{chouldechova2017fair,eubanks2018automating,benjamin2019race}. The process harm this paper theorizes is distinct and has a datafication dimension: when participation processes convert community witnessing into logged testimony and return no substantive response, they break the address-response structure through which community members sustain political agency \citep{oliver2004witnessing}. This harm is architecturally produced by documentation systems, not reducible to ordinary organizing fatigue. The category system that rendered community testimony legible as institutional input is the same system that forecloses acknowledgment that witnessing occurred. Naming it as harm requires governance frameworks to attend to participation processes themselves as sites of potential harm, not only to algorithmic outputs. An institution that recognized this harm would treat the absence of substantive response to community testimony as a governance failure requiring documented corrective action, distinct from the failure of the deployed system to perform equitably. The emotional labor burden empirical work documents \citep{10.1145/3772363.3790077,fox2017intersectional} is the empirical indicator of this structural injury. 

The third is the identification of governance architecture conditions from the counter-case. The conditions that enabled Oakland's partial escape from the legibility trap---\textit{binding authority}, \textit{external accountability}, \textit{burden reversal}, and \textit{ongoing accountability}---are empirical findings about what the cases with genuine accountability share that the cases without it do not \citep{corbett2023power,southerland2023master,irani2021oversight}. Future governance design work should treat these conditions as necessary constraints on institutional design. The four structural conditions Oakland built are individually reproducible without Oakland's specific organizing history. What they share is a portable principle: accountability authority must exist outside the governed institution before any specific participation event occurs. For policymakers, this means shifting the burden of proof to the agency, requiring public documentation of non-deployment demands and dissenting testimony, funding independent community review capacity, and making annual compliance review a condition of continued system use. For community organizations, the practical target is external leverage rather than better consultation: ordinance language, procurement conditions, records-access rights, and independent oversight bodies whose authority derives from law rather than from the goodwill of the institution being governed. CCOPS-model legislation circulated by civil liberties organizations provides a concrete template that compresses the organizing timeline significantly.

Arnstein's ladder already named what this paper demonstrates \citep{arnstein1969ladder}. The bottom rungs describe what institutions do to participation when they retain the authority to define it. The top rung requires authority that communities hold independently of the institution. The field has not yet theorized that structural condition as a governance design requirement. This theorization is the work participatory AI governance research must do to build counter-hegemonic dissent and subvert state extraction.

\section{Limitations and Future Work}

The corpus covers U.S. public sector cases with organized civil society and documented community opposition. The legibility trap argument predicts the mechanism operates in contexts with weaker civil society or more authoritarian state structures, but that prediction is untested. Where the liberal democratic assumption fails more completely, as in Aadhaar-linked welfare infrastructure in India or smart city surveillance governance in the Global South, the state constructs communities as administrative categories with no standing to oppose \citep{couldry2019costs,blili2024unsocial,benjamin2019race}. Future work should apply the CDA coding scheme to India's Facial Recognition System procurement, Aadhaar-linked benefit delivery governance, and EU AI Act impact assessments. The CDA was conducted by two coders working across all three cases, with codes compared and reconciled through deliberation. The absence markers category requires judgment about what communities said outside the official record, and future work should develop the coding scheme into a protocol with explicit decision rules, test it with additional coders across a larger corpus, and report reliability statistics alongside findings. A larger corpus would also enable computational narrative analysis of the discursive patterns this paper identifies qualitatively, potentially scaling the legibility trap analysis across the full range of CCOPS implementations in the twenty-five jurisdictions that had passed surveillance ordinances by 2021 \citep{irani2021oversight,southerland2023master}. The structural conditions this paper identifies were produced by specific historical circumstances in Oakland, organized resistance to surveillance infrastructure, the Domain Awareness Center fight as a galvanizing event, and the availability of legal expertise in Oakland Privacy as an organization. Specifying how those conditions can be cultivated elsewhere requires practitioner-facing research on how community organizations doing CCOPS-style governance work navigate specific institutional and legal constraints \citep{corbett2023power,maas2024beyond,delgado2022legal}.

\section*{Ethical Statement}
This research analyzes publicly available official government documents and publicly documented community testimony. No human subjects research was conducted. Ethical approval was not required for this study. All quoted documents are part of the public record and cited with full attribution. The analysis centers the documented testimony of community organizers, including Tawana Petty, whose public statements are cited here with the recognition that her labor and her documented exhaustion constitute evidence of the very harm this paper theorizes.

The authors declared no potential conflicts of interest with respect to the research, authorship, and publication of this article. No financial support was received for this research.

\section*{Acknowledgments}
The authors thank the research communities and community organizers whose public testimony constitutes the evidentiary core of this paper, including Tawana Petty and the Detroit Digital Justice Coalition, Oakland Privacy, and Secure Justice. We also thank the AIES 2026 reviewers whose comments substantially strengthened the paper. This work received no external financial support.

\bibliography{AIES26}

\end{document}